\documentclass[aps,prd,a4paper,twocolumn,showpacs,floats,superscriptaddress]{revtex4} 
\pdfoutput=1 
\usepackage{amsmath,amssymb}
\usepackage{placeins,float}
\usepackage{color,epsfig,rotate,graphicx}
\usepackage{hyperref} 

\def\beq{\begin{equation}}
\def\eeq{\end{equation}}
\def\bea{\begin{eqnarray}}
\def\eea{\end{eqnarray}}

\def\kms{~{\rm km s^{-1}}}
\def\mpc{{\rm Mpc^{-1}}}

\def\fig{{\rm Fig.}}
\def\bwt{\begin{widetext}}
\def\ewt{\end{widetext}}

\def\LCDM{$\Lambda${\rm CDM}}
\def\nin{\noindent} 

\begin{document}
\title{Clustering of quintessence on horizon scales\\ and its imprint on HI intensity mapping}
\author{Didam G.A. Duniya}
\affiliation{Physics Department, University of the Western Cape, Cape Town 7535, South Africa}

\author{Daniele Bertacca}
\affiliation{Physics Department, University of the Western Cape, Cape Town 7535, South Africa}

\author{Roy Maartens}
\affiliation{Physics Department, University of the Western Cape, Cape Town 7535, South Africa}
\affiliation{Institute of Cosmology \& Gravitation, University of Portsmouth, Portsmouth PO1 3FX, UK }
\date{\today}
\begin{abstract}\nin
Quintessence can cluster only on horizon scales. What is the effect on the observed matter distribution? To answer this, we need a relativistic approach that goes beyond the standard Newtonian calculation and deals properly with large scales. Such an approach has recently been developed for the case when dark energy is vacuum energy, which does not cluster at all. We extend this relativistic analysis to deal with dynamical dark energy. Using three quintessence potentials as  examples, we compute the  angular power spectrum for the case of an HI intensity map survey. Compared to the concordance model with the same small-scale power at $z=0$, quintessence boosts the angular power by up to ${\sim} 15\%$ at high redshifts, while power in the two models converges at low redshifts. The difference is mainly due to the background evolution, driven mostly by the normalization of the power spectrum today. The dark energy perturbations make only a small contribution on the largest scales, and a negligible contribution on smaller scales. Ironically, the dark energy perturbations remove the false boost of large-scale power that arises if we impose the (unphysical) assumption that the dark energy is smooth.   
\end{abstract}  
\pacs{95.36.+x, 98.62.Py, 95.35.+d}
\maketitle
\section{Introduction}\label{sec:intro}%
Upcoming galaxy surveys in the optical/infrared and in the 21cm emission of neutral hydrogen (HI) will extend to higher redshifts and wider areas of the sky, covering greater volumes and including scales approaching and larger than the Hubble radius $H^{-1}(z)$. On these scales, the usual Newtonian approach is inadequate. A fully relativistic analysis is necessary if we are to extract maximal and accurate information from large-volume galaxy surveys~\cite{Yoo:2010jd}--\cite{Yoo:2013tc}. Relativistic effects become significant in the same large-scale regime that is crucial for:\\ (a)~testing of dark energy models, modified gravity models and general relativity itself;\\ (b)~measuring the signal of primordial non-Gaussianity in large-scale structure. 

In particular, the Planck data has constrained local non-Gaussianity to be small \cite{Ade:2013ydc}, 
\beq
f_{NL}=2.7 \pm 5.8, 
\eeq
so that an accurate measurement of $f_{NL}$ in large-scale structure will require careful accounting of the relativistic effects in the observed overdensity \cite{Bruni:2011ta}--\cite{Yoo:2011zc}, \cite{Maartens:2012rh}. The key problem with large-scale correlations is cosmic variance, but this can be overcome if we have multiple tracers of the underlying matter distribution \cite{Yoo:2011zc,Abramo:2013awa}. 

Thusfar, the relativistic effects in the observed overdensity (or the fractional HI brightness temperature fluctuations) have been computed for flat \LCDM\, models (except for \cite{Hall:2012wd,Lombriser:2013aj}). 
If dark energy is dynamical, e.g. a quintessence model (QCDM), then it can cluster on super-Hubble scales, unlike the vacuum energy $\Lambda$ which does not cluster at all.  
Here we extend the analysis of relativistic effects in the observed overdensity from \LCDM\,  to the case of dynamical dark energy. The  clustering of dark energy on horizon scales should have some effect on the matter power, and to accurately identify this effect, a fully relativistic analysis is needed. We illustrate this via three different QCDM potentials -- Ratra-Peebles, supergravity and double exponential  -- 
whose dark energy equation of state cannot be approximated by a constant or a simple parametrization (see Appendix A for details of these models). We compute the angular power spectrum for an HI intensity mapping survey, comparing it to a \LCDM\, model. 

\begin{figure}\centering 
\includegraphics[scale=0.4]{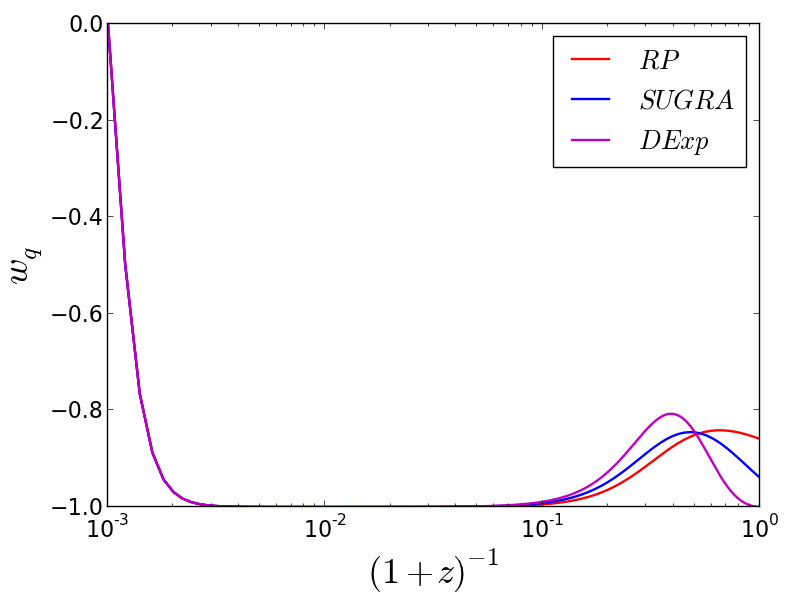} \\
\includegraphics[scale=0.4]{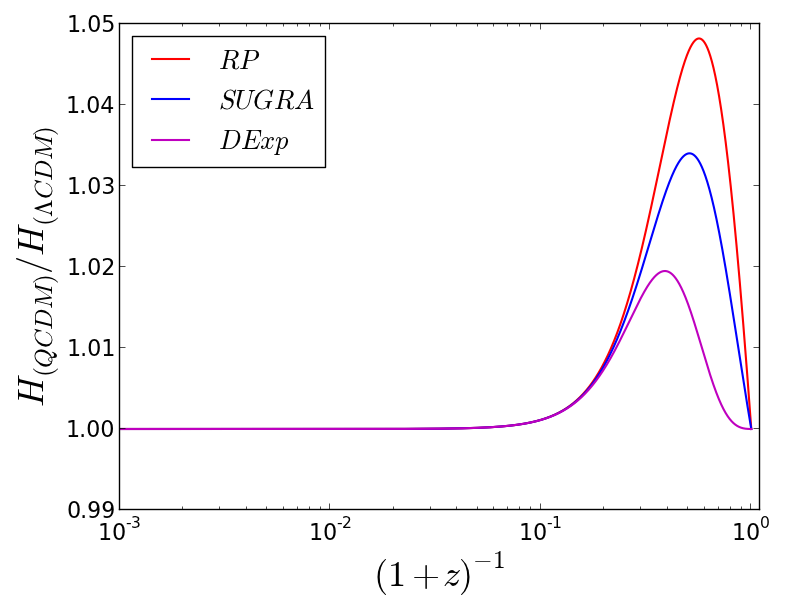} \\
\includegraphics[scale=0.4]{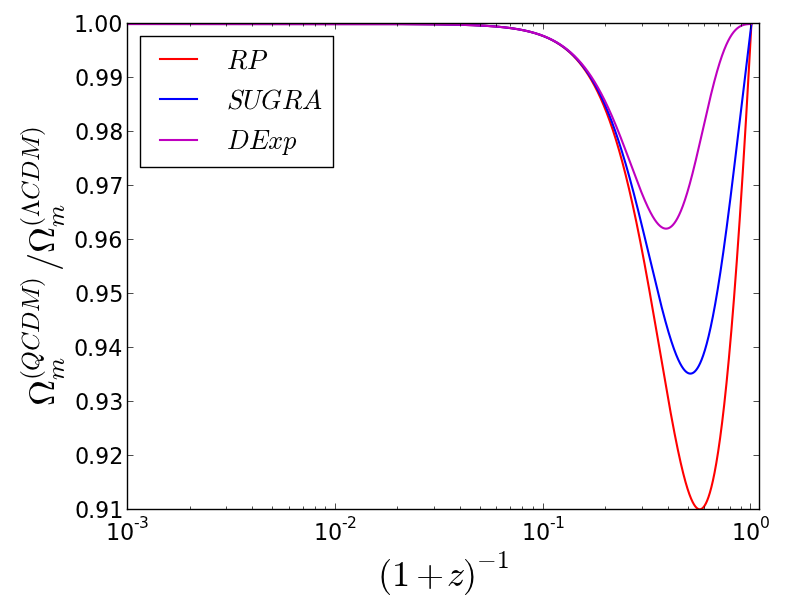}  
\caption{The cosmic evolution of the QCDM equation of state parameter ({\em top}), Hubble rate ({\em middle}) and matter density parameter ({\em bottom}).}\label{fig:1}
\end{figure}

Our aim is not to fit the QCDM models to the data, but to show how these models differ from \LCDM. Therefore we choose parameters and initial conditions so that the QCDM models have the same $\Omega_{m0}$ and $H_0$ as the \LCDM\, model. In particular, this normalizes the matter power spectra to be the same on small (linear) scales at $z=0$, thus isolating differences on large scales. The evolution of the equation of state parameters, the Hubble rate and the matter density for the 3 models, is shown in Fig. \ref{fig:1}. This background evolution is driven by our normalization of the power spectra.  Note the deviation of the equation of state from $-1$ for $z\lesssim 10$ in these models.

\section{Quintessence: relativistic perturbations}\label{sec:quintGR} %

For a general quintessence field  $\varphi$ driven by a potential $U(\varphi)$, the Friedmann and acceleration equations are 
\bea\label{eq:gen:1}
\mathcal{H}^{2} &=& \frac{8{\pi}Ga^{2}}{3}\left[\rho_m +\frac{{\varphi}'^{2}}{2a^{2}} + U(\varphi)\right] \\
&=& \mathcal{H}^{2}(\Omega_m+\Omega_q), \nonumber
\\ \label{eq:gen:2}
{\cal H}' &= &-{1\over 2}{\cal H}^{2}\left(1+3w_q\Omega_q\right),\quad w_q = \frac{\varphi'^2-2a^2U} {\varphi'^2+2a^2U}. 
\eea
Quintessence evolves under the Klein-Gordon equation,
\beq\label{eq:gen:3}
{\varphi}'' +2{\cal H}\varphi' +a^2\frac{\partial U}{\partial\varphi} = 0,
\eeq%
and its equation of state is governed by
\beq\label{eq:gen:4}
w_{q}'=-3{\cal H}(1-w^{2}_{q})-(1-w_{q}){{\varphi}'\over U}\frac{\partial U }{\partial\varphi}. 
\eeq%

The perturbed metric in longitudinal gauge is
\beq\label{eq:gen:5}%
ds^2 = a^2\left[-(1+2\Phi)d{\eta}^2 + (1-2\Phi)d\vec{x}\,^2\right],
\eeq
and the gravitational potential $\Phi$  is given by the relativistic Poisson equation
\beq\label{eq:gen:6} 
\nabla^2\Phi={3\over 2}{\cal H}^2(\Omega_m\Delta_m+\Omega_q\Delta_q),
\eeq
where the {\em comoving} density contrasts are defined by
\beq\label{eq:gen:6b}
\Delta \equiv \delta-3(1+w){\cal H} V,~~u^\mu=a^{-1}(1-\Phi,\partial^iV).
\eeq
Here $u^\mu$ is the $4$-velocity, and $V$ is the peculiar velocity potential. Note that \eqref{eq:gen:6b} applies to each species, i.e. the comoving density contrasts are in the comoving gauge of each species separately.

On sub-Hubble scales, we can neglect $\Delta_q$ and $\Delta_m \approx \delta_m$,  leading to the Newtonian Poisson equation that is typically used in large-scale structure analysis: $\nabla^2\Phi=(3/2){\cal H}^2\Omega_m\delta_m$. But on large scales $\delta_m$ is no longer an accurate tracer of the potential and the relativistic Poisson equation \eqref{eq:gen:6} must be used. 
The relativistic Poisson equation shows that dark energy clustering (on large scales) enhances the potential for a given matter overdensity. Alternatively, to attain a given magnitude of potential, one requires less matter clustering when the dark energy clusters. 

The potential evolution is driven by the total momentum density, 
\beq\label{eq:gen:7}
{\Phi}' +{\cal H}\Phi = -{3\over 2}{\cal H}^2\left[\Omega_m V_m +(1+w_q)\Omega_q V_{q}\right].
\eeq%
Matter fluctuations obey energy-momentum conservation:
\bea\label{eq:gen:8}
{\Delta}'_{m} - {9\over 2}{\cal H}^{2}\Omega_q(1+w_q)\left( V_{m}-V_{q}\right) &=& -\nabla^{2}V_{m}, \\
\label{eq:gen:9}
V'_{m} + {\cal H}V_{m} &=& -{\Phi},
\eea%
and for quintessence,
\bea
{\Delta}'_{q} - 3w_{q}{\cal H}\Delta_{q} &-& \frac{9}{2}{\cal H}^{2}\Omega_m(1+w_{q})\left(V_{q}-V_m\right)  \nonumber\\ \label{eq:gen:10}
 & =& -(1+w_{q})\nabla^{2}V_{q},\\ \label{eq:gen:11}
 V'_{q} + {\cal H}V_{q}& =& -\frac{c_{sq}^2}{(1+w_q)}\Delta_q-\Phi ,
\eea
where $c_{sq}$ is the dark energy sound speed. 

The equations \eqref{eq:gen:6}--\eqref{eq:gen:11} hold for any form of dark energy: one simply needs to specify $w_q(a)$ and $c_{sq}(a)$. For example, fluid models of dark energy specify these parameters ad hoc. In the case of quintessence, these parameters are self-consistently determined: $w_q$ is determined by the potential $U(\varphi)$ via \eqref{eq:gen:3}--\eqref{eq:gen:4}, and for any $U(\varphi)$, the physical sound speed of quintessence is $c_{sq}=1$.

\subsection{Power spectrum on large scales}

\begin{figure}\centering 
\includegraphics[scale=0.4]{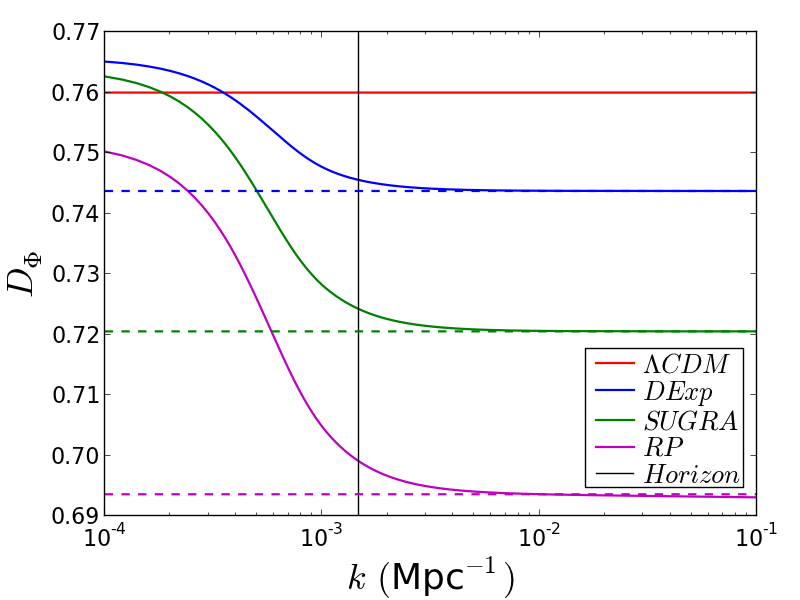} \\ %
\includegraphics[scale=0.4]{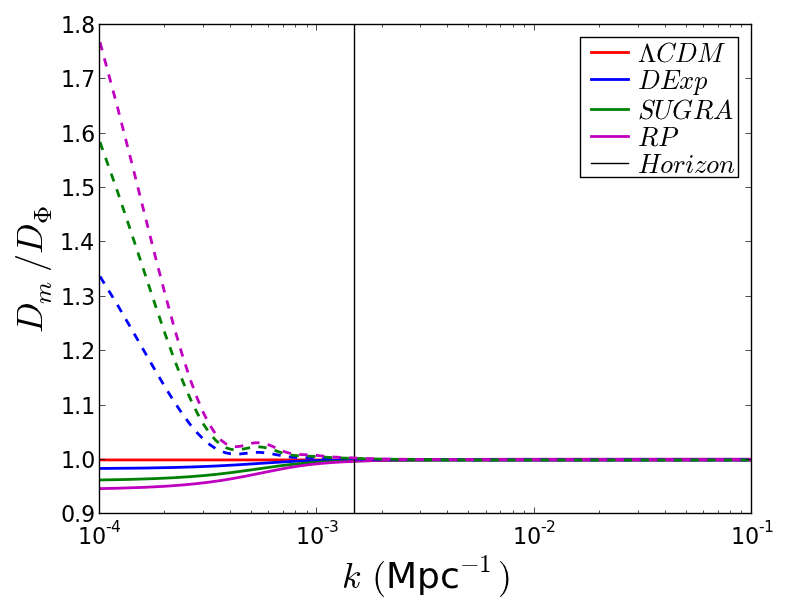} 
\includegraphics[scale=0.4]{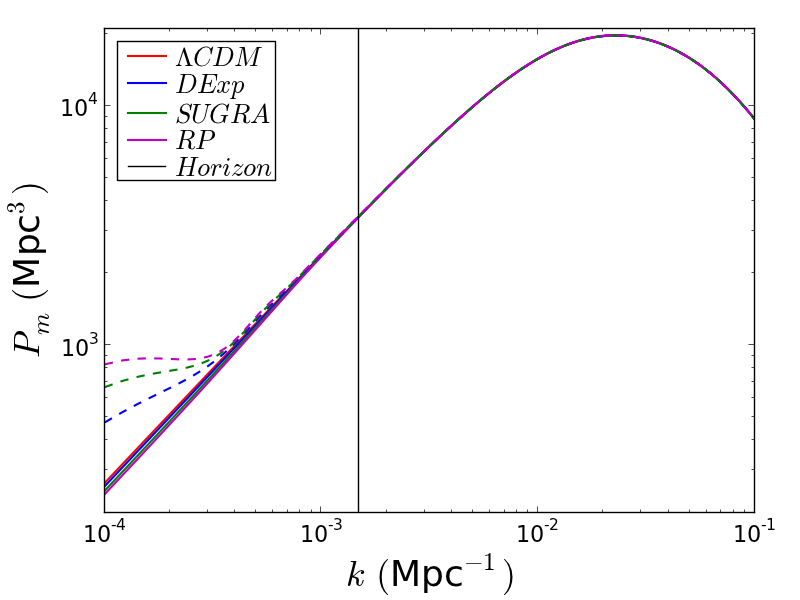} \\%
\includegraphics[scale=0.4]{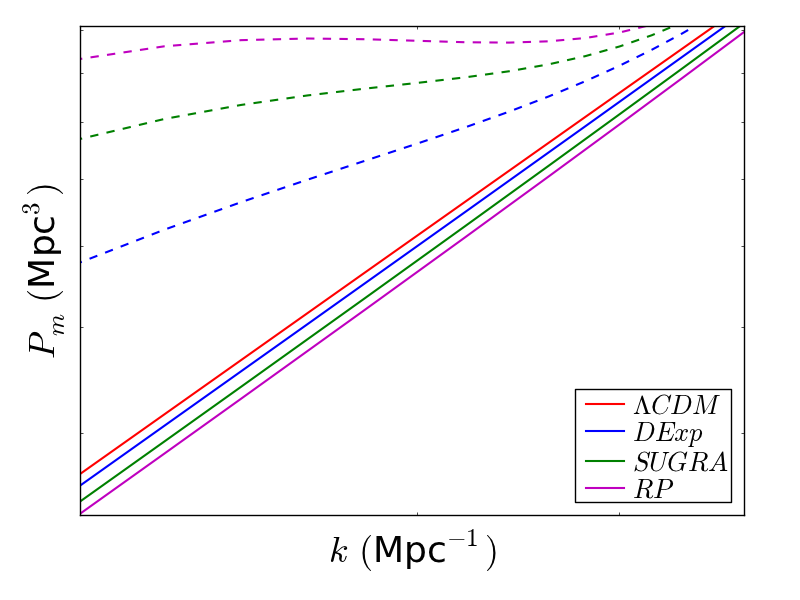}
\caption{For the QCDM models, from top to bottom: potential growth function $D_{\Phi}(k,0)$; matter power factor $D_m(k,0)/D_\Phi(k,0)$; matter power spectrum $P_{m}(k,0)$, with zoom-in at large scales. The dashed lines give the (unphysical) case of smooth quintessence (i.e. perturbations forced to zero).
}\label{fig:2}
\end{figure} %

We initialize the integrations at $a_d=(1+z_d)^{-1}=10^{-3}$ (soon after decoupling), neglecting photons and neutrinos. The initial background densities of matter and quintessence are determined by the requirement that $\Omega_{m0}$ and $H_0$ match the \LCDM\, values (we take $\Omega_{m0} = 0.27$, $H_0 = 70.8\kms\mpc$). Quintessence is assumed to be initially in a tracking regime, $w_q(z_d)=0$. The effect of the radiation era on perturbations is modelled via a transfer function $T(k)$, which is the same in QCDM and \LCDM, since dark energy plays a negligible role in the radiation era. Since $\Omega_q(z_d)\ll 1$, we use the Einstein de Sitter initial condition  $\Phi'(z_d)=0$. The adiabatic initial conditions
\beq \label{adic}
S_{qm}(z_d)=0=S'_{qm}(z_d), ~~~~ S_{qm}={\delta_q \over (1+w_q)}-\delta_m,
\eeq
together with the Poisson and other perturbation equations, determine the initial matter and quintessence fluctuations (see Appendix B). 

We connect the primordial inflationary perturbations to the decoupling epoch via the potential:
\bea\label{eq:lps:1} 
\Phi(k,z) &=& (1+z){D_{\Phi}(k,z)}\,\Phi_d(k),\\ \label{phid}
\Phi_d(k) &=& \frac{9}{10}\Phi_p(k) T(k),\\ \label{eq:lps:4}
\Phi_p(k) &=&   A\frac{\Omega_{m0}}{D_{\Phi }(k,0)}\, \left(\!\frac{k}{H_0}\!\right)^{(n-4)/2}.
\eea
Here $(1+z)D_\Phi$ is the potential suppression function, normalized by  $(1+z_d)D_{\Phi}(k,z_d)=1$.   The constant $A$ gives the primordial amplitude of curvature perturbations, and $n=0.96$ is the scalar spectral index.
For the matter overdensity, 
\bea
\label{eq:lps:3}
\Delta_m(k,z) &=& -\frac{2}{3\Omega_{m0}}{k^2 \over H_0^2}D_m(k,z)\Phi_d(k),
\eea
where the matter growth function is given by
\beq\label{eq:Dm}
{D_m(k,z) \over D_m(k,z_d)} ={\Delta_m(k,z) \over  \Delta_m(k,z_d)}. 
\eeq
In \LCDM, we have 
\beq
\mbox{\LCDM:}~~D_\Phi(k,z)=D_m(k,z) \equiv D(z). \label{dlcdm}
\eeq
For QCDM this is no longer true. However, the normalization we have chosen means that \eqref{dlcdm} does hold at $z=0$ on small scales where the quintessence perturbations are negligible:
\beq \label{pmn}
\mbox{QCDM:}~~D_\Phi(k,0)=D_m(k,0) ~~\mbox{for}~ k\gg H_0.
\eeq
This property is shown in Fig.~\ref{fig:2}.

In order to incorporate the largest scales, we use not the longitudinal density contrast $\delta_m$, but the comoving density contrast $\Delta_m$, for the matter power spectrum: $P_m=P_{\Delta_m}$. By \eqref{eq:lps:1}--\eqref{eq:lps:3}, the matter power spectrum is
\beq
P_m(k,z) ={9A^2 \over 50 \pi^3H_0^n} k^nT^2(k) \left[\frac{D_m(k,z)}{D_{\Phi }(k,0)} \right]^2.
\eeq
From \eqref{pmn}, we see that the power spectrum for QCDM today will match that of \LCDM\ on small scales. The changes induced by quintessence will be imprinted in $P_m$ via $D_m/D_{\Phi 0}$ and will show up on the largest scales. These features are illustrated in 
Fig.~\ref{fig:2}, which shows the growth functions and power spectra for the QCDM models. 

Quintessence clusters on large scales, leading to a scale-dependent $D_\Phi$ on large scales. On small scales, $D_\Phi$ becomes scale-independent, since quintessence does not cluster on these scales. The offset between this constant value and that of \LCDM\ follows from the different expansion history of quintessence that is necessary to achieve the same $\Omega_{m0}$.  

We also show the unphysical case of smooth quintessence, i.e. forcing quintessence perturbations to be zero,
$V_q = \delta_q = \Delta_q = 0$, and ignoring the evolution equations \eqref{eq:gen:10}--\eqref{eq:gen:11}. This approximation is often made -- but it clearly breaks down badly on large scales. The smooth quintessence leads to an inconsistency in the perturbations equations, so that the results are gauge-dependent. In longitudinal gauge, smooth quintessence predicts a  strong boost of matter power on the largest scales -- whereas in fact the quintessence perturbations act to slightly {\em suppress} the matter power on these scales, when the small-scale power is normalized to the \LCDM\, power at $z = 0$. 

These results are consistent with previous work \cite{Unnikrishnan:2008qe}--\cite{Jassal:2012pd}.  In the next section we extend previous results to consider the observable angular power spectrum $C_\ell(z)$, including all relativistic effects.

\begin{figure}\centering 
\includegraphics[scale=0.45]{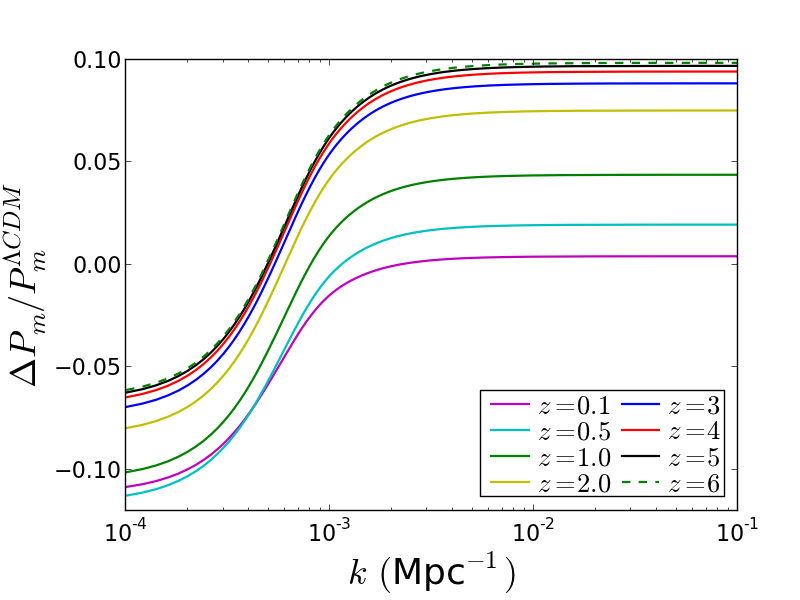}    %
\caption{Fractional change in the matter power spectrum for inhomogeneous QCDM (Ratra-Peebles potential), relative to \LCDM\, at various redshifts $z$ (here $\Delta P_m \equiv P^{\rm QCDM}_m - P^{\Lambda{\rm CDM}}_m$).} \label{fig:3} 
\end{figure}

Owing to our normalization at $z=0$, the QCDM power spectrum today matches that of \LCDM\, on small scales. This is not true for $z>0$, as illustrated in Fig.~\ref{fig:3}. In QCDM, the matter power is enhanced on sub-Hubble scales with increasing $z$. This is necessary in order to achieve the same power as \LCDM\, today -- given that in QCDM, $\Omega_m(z)$ is smaller and $H(z)$ is bigger, for $0<z \lesssim 10$. On super-Hubble scales, power is suppressed, more strongly at low redshifts, at the level of ${\sim} 5-10\%$ for the Ratra-Peebles potential. 

\subsection{Relativistic effects in the observed overdensity }\label{sec:Cls} %

Up to now we have worked with the matter density contrast $\Delta_m$ and its power spectrum $P_m$, i.e. with the perturbations in the total matter, dark and baryonic. However, we cannot observe the total matter fluctuations, only tracers of them, such as galaxies. 
The galaxy density contrast is 
\beq \label{bias}
\Delta_g(k,z)=b(z)\Delta_m(k,z),
\eeq
where $b$ is the scale-independent bias on linear scales (and for Gaussian primordial perturbations).
The relation $\delta_g=b\delta_m$ is gauge-dependent on very large scales and thus cannot be used to define the physical bias \cite{Challinor:2011bk,Bruni:2011ta}. Since the bias is physically defined in the matter rest frame, we must use the comoving density perturbation in the bias definition \eqref{bias}.

However, there is a problem with using $\Delta_g$ as a probe of matter density perturbations: it does not correspond to the observed overdensity  on large scales, because of relativistic effects -- lightcone and redshift effects -- in the observational process:\\ (1) Firstly, we need to incorporate the well-known effects of redshift space distortions and weak lensing. These are typically the dominant relativistic effects in the observed overdensity.\\ (2) Secondly, there are also other relativistic effects that arise when we correctly define the physical, observed galaxy overdensity $\Delta^{\rm obs}_g$ \cite{Yoo:2010jd,Yoo:2010ni,Bonvin:2011bg,Challinor:2011bk,Jeong:2011as}. 

The observed overdensity is unique and physically defined. It does not correspond to any of the standard gauge-invariant definitions of overdensity -- but it is automatically gauge-invariant. We can write it in the form
\bea\label{delta_g} 
\Delta^{\rm obs}_g( {\bf n},z) &=&  {\Delta}_g ( {\bf n},z)+ {\Delta}_{\rm loc}( {\bf n},z)+ {\Delta}_{\rm int}( {\bf n},z),
\eea
where $z$ is the observed redshift (with corresponding background scale factor $a=(1+z)^{-1}$) and  {\bf n} is the unit direction of observation. The comoving position of the source is ${\bf x}=r(z){\bf n}$, where the comoving radial distance is
\bea\label{eq:cls:13}
r(z) = \int^{z}_{0}{ \frac{d\tilde{z}}{(1+\tilde{z}) {\cal H}(\tilde{z})} }.
\eea
In the gauge we have chosen, the local and integrated terms are
\bea \label{delloc}
{\Delta}_{\rm loc} &=&  -\frac{1}{\mathcal{H}} (n^i\partial_i)^2 V_m + \Delta_{\rm loc}^{\rm rel} ,\\
 \label{delint}
{\Delta}_{\rm int}&=& 
-\left(1-Q\right)  \int_0^{r} {d\tilde r} \left(r-\tilde r\right) {r \over\tilde r} \nabla^2_\perp \Phi  + \Delta_{\rm int}^{\rm rel}.
\eea 
The first term in \eqref{delloc} is the standard redshift space distortion term and the first term in \eqref{delint} is the standard weak lensing term. Both of these standard terms are relativistic corrections to a Newtonian approach. Here $Q(z)$ is the magnification bias \cite{Jeong:2011as}, and $\nabla^2_\perp= \nabla^2-(n^i\partial_i)^2-2r^{-1}n^i\partial_i$ is the screen-space Laplacian.

The additional local and integrated relativistic correction terms are \cite{Jeong:2011as}
\bea 
{\Delta}_{\rm loc}^{\rm rel} &=&  \left(3-b_e\right)\mathcal{H} V_m \nonumber\\
&&{}+\left[b_e -2Q-\frac{\mathcal{H}'}{\mathcal{H}^2}-2\frac{\left(1-Q\right)}{r \mathcal{H}}\right] n^i \partial_i V_m ~~~~~~~~~~ \nonumber \\
&&{}+\left[4Q -b_e -1 +\frac{\mathcal{H}'}{\mathcal{H}^2} +2\frac{\left(1-Q\right)}{r \mathcal{H}}\right]\Phi  \nonumber \\
&&{} + {1\over \mathcal{H}} \Phi', \label{locgr}\\ \nonumber\\
{\Delta}_{\rm int}^{\rm rel}&=& 4  \frac{\left(1-Q\right) }{r}\int_0^{r} {d\tilde r}\, \Phi \nonumber\\
&&{}- 2 \left[b_e -2Q-\frac{\mathcal{H}'}{\mathcal{H}^2}-2\frac{\left(1-Q\right)}{r \mathcal{H}}\right] \int_0^{r} {d\tilde r}\, \Phi'.
\label{intgr}
\eea 
Here the galaxy evolution bias is
\bea\label{eb}
b_{e}(z) =- \frac{d\ln N_g}{d\ln(1+z)}, 
\eea
where $N_g(z)$ is the comoving number density.

\section{Observed angular power spectrum}

We could use $P_g^{\rm obs}\equiv P_{\Delta_g^{\rm obs}}$ as a measure of the observed galaxy overdensity (see \cite{Lombriser:2013aj}), but it is not directly observable, since it describes fluctuations on a constant time slice rather than on the observed past lightcone. It is useful to use instead the observed angular power at different redshifts, $C_\ell(z)$. 
 (See \cite{Bertacca:2012tp} for an alternative approach based on the observed correlation function $\xi({\bf n}_1,z_1, {\bf n}_2,z_2)$.)
Note also that we avoid the flat-sky approximation by using $C_\ell(z)$. This is also important for consistently dealing with large scales.

We expand in spherical harmonics
\beq
\Delta^{\rm obs}_g( {\bf n},z)=\sum_{\ell m} a_{\ell m}(z) Y_{\ell m}({\bf n}),
\eeq
to get the angular power spectrum at a given redshift:
\bea\label{eq:gr:7} 
C_{\ell}(z)& =& \left\langle \left| a_{\ell m}(z)\right|^2 \right\rangle, \\
a_{\ell m}(z) &=& \int d^2{\bf n}\, Y^*_{\ell m}({\bf n}) \Delta^{\rm obs}_g( {\bf n},z).
\eea 
Thus $C_\ell(z)$ involves the auto- and cross-correlations of all the terms on the right of \eqref{delta_g}, using \eqref{delloc}--\eqref{intgr}. For example,
\bea \label{intcl}
\left\langle \left| \Delta_g( {\bf n},z) \right|^2\right\rangle_\ell &\propto &{b^2(z) \over (\Omega_{m0}H_0^2)^2} \times \\
&&{} \int dk \,k^6P_{\Phi_d}(k)D_m^2( k,z) [j_\ell(r(z)k)]^2. \nonumber
\eea

\begin{figure}\centering 
\includegraphics[scale=0.38]{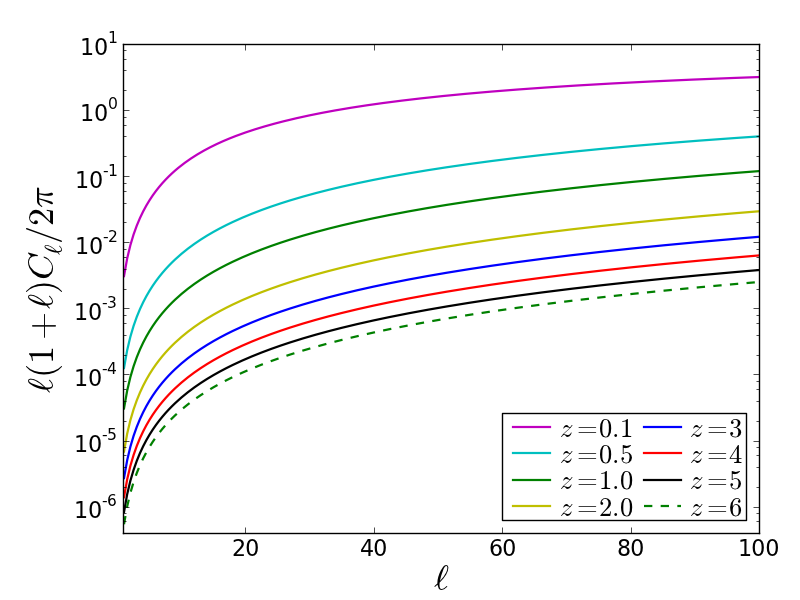} \\%
\includegraphics[scale=0.374]{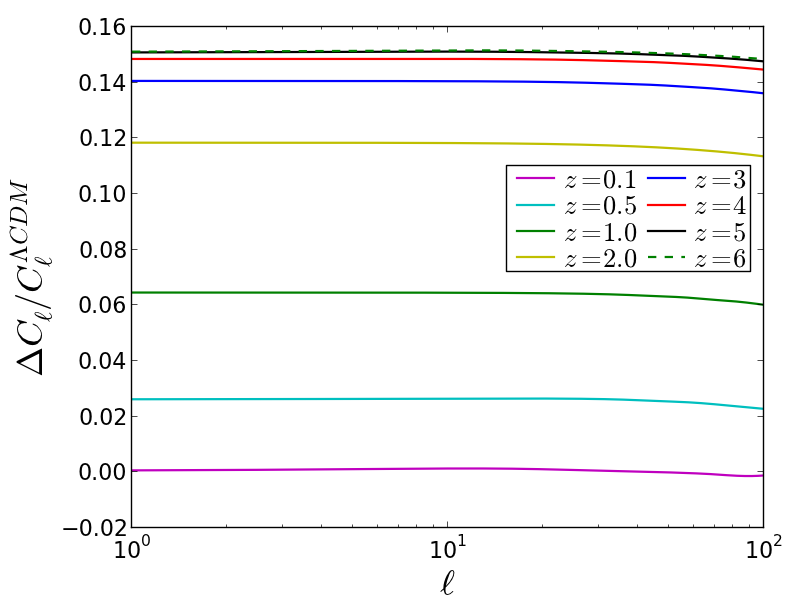} \\%
\includegraphics[scale=0.38]{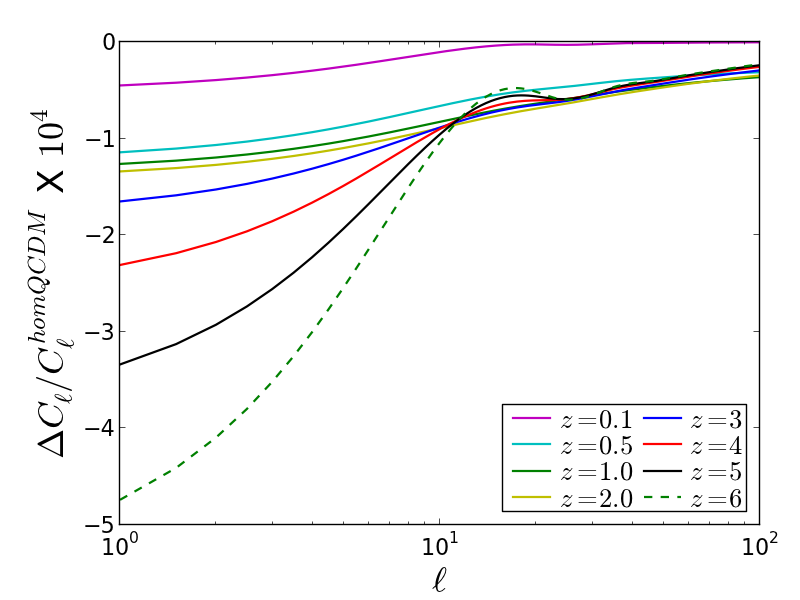} %
\caption{The angular power spectrum of the fractional HI brightness temperature fluctuations for QCDM (Ratra-Peebles potential) at various redshifts $z$ ({\em top}). The fractional change in angular power spectrum between: QCDM and \LCDM, where $\Delta C_{\ell}\equiv C^{\rm QCDM}_{\ell} - C^{\Lambda{\rm CDM}}_{\ell}$ ({\em middle});  inhomogeneous and homogeneous QCDM, where $\Delta C_{\ell}\equiv C^{\rm QCDM}_{\ell} - C^{\rm homQCDM}_{\ell}$ ({\em bottom}).}\label{fig:4} 
\end{figure} %

\subsection{HI intensity map survey}

For an intensity map in HI, the observed brightness temperature fractional perturbations are given by \cite{Hall:2012wd}
\beq \label{h1btf}
\Delta^{\rm obs}_{T_b}= \Delta^{\rm obs}_g- 2 {\delta d_L \over d_L},
\eeq
where $d_L$ is the luminosity distance. The relativistic expression for $\delta d_L$ contains the weak lensing and other relativistic terms above -- and the result is that
\beq
\Delta^{\rm obs}_{T_b}= \Delta^{\rm obs}_g\Big|_{Q=1}.
\eeq
In other words, the brightness temperature fractional perturbations in an HI intensity map are given by the galaxy overdensity in the case when the magnification bias is $Q=1$. The most important consequence is that the weak lensing term cancels out in this case, and we get
\bea 
\Delta^{\rm obs}_{T_b} &=& {\Delta}_g ( {\bf n},z)-\frac{1}{\mathcal{H}} (n^i\partial_i)^2 V_m   
\nonumber \\ &&{}  +3\mathcal{H} V_m -\left(2+\frac{\mathcal{H}'}{\mathcal{H}^2}\right) n^i \partial_i V_m
\nonumber \\ &&{}  +\left(3+\frac{\mathcal{H}'}{\mathcal{H}^2}\right)\Phi + {1\over \mathcal{H}} \Phi' \nonumber\\
&&{}+2 \left( 2+\frac{\mathcal{H}'}{\mathcal{H}^2}\right) \int_0^{r} {d\tilde r}\, \Phi'. \label{h1btf2}
\eea 
For simplicity, we used $N_g(z)=\,$const, so that $b_e=0$. The first line has the Kaiser distortion term. The second line has relativistic velocity corrections. Relativistic potential terms are in the third line (local terms) and fourth line (integrated).

Figure~\ref{fig:4} shows the angular power spectrum of the {\em fractional} (i.e. dimensionless) HI brightness temperature fluctuations, 
\bea\label{angh1} 
C^{T_b}_{\ell}(z) &=& \left\langle \left| a_{\ell m}(z)\right|^2 \right\rangle, \\
a_{\ell m}(z) &=& \int d^2{\bf n}\, Y^*_{\ell m}({\bf n}) \Delta^{\rm obs}_{T_b}( {\bf n},z),
\eea 
for a QCDM model at various redshifts (top panel). The integrated term in \eqref{h1btf2} makes a negligible contribution \cite{Hall:2012wd}, and the dominant term is the redshift space distortion, as pointed out in \cite{Hall:2012wd}.  

The fractional difference between QCDM and \LCDM\, (middle panel) reflects the effect of the equation of state parameter $w_{q}$  on the matter power spectrum of \fig~\ref{fig:3}. This fractional difference increases with redshift and asymptotes to a maximum that is determined by our normalization. Note that the strongest growth in the fractional difference occurs for $1\lesssim z\lesssim 2$, where the deviation of $w_q$ from $-1$ is largest (\fig~\ref{fig:1}). However, we do not observe any significant scale-dependence in the changes of the $C_\ell$, compared with the matter power spectrum case in \fig~\ref{fig:3}. This is due to the fact that the large-scale changes in $P_m(k,z)$ are suppressed by the integration over $k$ to get $C_\ell(z)$ (see for example \eqref{intcl}). In other words, the $\sim 5-15\%$ suppression of power in $P_m$ is reduced to sub-percent in the $C_\ell$.

The bottom panel of \fig~\ref{fig:4} shows the small changes in the angular power spectrum of the (dimensionless) HI brightness temperature fluctuations between the physical model of clustering quintessence and the unphysical smooth quintessence. On large scales, smooth quintessence provides a false boost in angular power $C_\ell(z)$, which grows as $z$ increases. This is consistent with the behaviour of $P_m(k,z)$ in Fig. \ref{fig:2}.

\section{Conclusion}\label{sec:concl} %

We have provided a fully relativistic treatment of dark energy and matter perturbations in the post-recombination universe. This requires a careful accounting of large-scale modes (and thus of gauge issues), and we have taken care to use the relativistic Poisson equation and to define the bias in terms of the comoving matter overdensities. This is important for deriving the correct matter power spectrum in the presence of inhomogeneous dark energy. Furthermore, we need to incorporate all lightcone and redshift effects by defining the observed overdensity. Our equations are a simple generalization of previous work on relativistic effects in the observed overdensity from \LCDM\, to dynamical dark energy models (see also \cite{Lombriser:2013aj}).

We illustrated the implications of the generalized equations for three QCDM models, each of which has nontrivial evolution in the quintessence equation of state $w_q$ at low redshifts, so that a simple parametrization of $w_q$ is not practical. In order to isolate the effect of quintessence on large-scale galaxy overdensity, we normalized the QCDM power spectrum today so that it agrees on sub-Hubble scales with the \LCDM\, power spectrum. We find that super-Hubble clustering has a very small effect on the large-scale power spectrum today. However, it is essential to incorporate the large-scale quintessence perturbations: the unphysical assumption of smoothness of quintessence (i.e. forcing the perturbations to vanish) violates the Einstein equations and leads to a false boost of large-scale power. The self-consistent treatment of inhomogeneous quintessence shows a small suppression of large-scale power today.

For higher redshifts, there is a boost of matter power on small scales. However this is mainly due to background evolution effects -- since a stronger growth of matter perturbations is needed in the presence of a lower $\Omega_m(z)$ and higher $H(z)$, in order to arrive at the \LCDM\, value of $\Omega_{m0}$.

We considered the case of an HI intensity map survey, where the observed fractional brightness temperature fluctuation is given by the observed galaxy (number) overdensity in the case of magnification bias $Q=1$, which removes the weak lensing convergence contribution. We computed the angular power spectra $C_\ell(z)$ for a QCDM model, showing that the large-scale suppression of power is further reduced relative to $P_m(k,z)$.

\[ \]
{\bf Acknowledgements:}
This work was supported by the South African Square Kilometre Array Project, the South African National Research Foundation and by a Royal Society (UK)/ National Research Foundation (SA) exchange grant. RM was also supported by the UK Science \& Technology Facilities Council (grant nos. ST/H002774/1 and ST/K0090X/1).

\appendix 

\section{QCDM models}

The Ratra-Peebles (RP) potential is \cite{Alimi:2009zk,Copeland:2006wr,BarroCalvo:2006wd}  
\begin{equation}\label{eq:4:28}
U(\varphi)=\frac{M^{4+\alpha}}{\varphi^{\alpha}},\quad \alpha > 0.
\end{equation}
The mass scale $M$ is chosen so that $(\sqrt{8{\pi}G})^{2+\alpha}M^{4+\alpha} / (3H^2_0) = 0.58$ \cite{BarroCalvo:2006wd} and $\alpha \simeq 0.501898922$. We obtain $w_{q0} \simeq -0.85927$.

The SUGRA potential \cite{Brax:1999gp,Brax:2000yb, Copeland:2006wr, Alimi:2009zk} is a super-gravity correction to the RP potential:
\begin{align}
U(\varphi) =\; \frac{M^{4+\alpha}}{\varphi^{\alpha}}\exp \left({4{\pi}G}\varphi^2\right),
\end{align}%
where we choose $(\sqrt{8{\pi}G})^{2+\alpha}M^{4+\alpha} / (3H^2_0) = 0.45$ \cite{BarroCalvo:2006wd} and $\alpha \simeq 0.65705469$. This leads to $w_{q0} \simeq -0.9395$.

The double exponential (DExp) potential is~\cite{Barreiro:1999zs,Brax:2000yb,BarroCalvo:2006wd,Bassett:2007aw}
\beq\label{eq:dexp:1}
U(\varphi)=M^{4}_{1}\exp \left(\alpha\varphi\sqrt{8\pi G} \right)+M^{4}_{2}\exp \left(\beta\varphi\sqrt{8\pi G} \right).
\eeq%
The behaviour of $\Omega_q$ and $w_q$ is very sensitive to the choice of the parameters $\alpha$ and $\beta$. We choose $M_1$ via $8{\pi}GM^4_1 / (3H^2_0) = 0.88926578$ \cite{BarroCalvo:2006wd}, set $M^4_2 = 0.4101M^4_1$ and take $\beta = 1$. Then we calculate $\alpha \simeq -6.25166029$ and find $w_{q 0} \simeq -0.99993$.

\begin{widetext}

\section{Adiabatic initial conditions} 

The adiabatic initial conditions
\eqref{adic}, together with $\Phi'(z_d)=0$,  lead to the initial values
\bea
\Delta_m &=& \frac{-2k^2\left[k^2 + 9{\cal H}^2(c^2_{sq}-c^2_{aq})\right]}{3{\cal H}^2\left[k^2(1+w_q-w_q\Omega_m) + 9{\cal H}^2\Omega_m(c^2_{sq}-c^2_{aq})\right]}\,\Phi_d(k),\\
\Delta_q &=& \frac{-2k^4(1 + w_q)}{3{\cal H}^2\left[k^2(1+w_q-w_q\Omega_m) + 9{\cal H}^2\Omega_m(c^2_{sq}-c^2_{aq})\right]}\, \Phi_d(k), \\
V_m &=& \frac{-2}{3{\cal H}}\left[\frac{k^2(1+w_q-w_q\Omega_m)(1 -3c^2_{sq}+3c^2_{aq}) + 3\Omega_m(k^2 + 3{\cal H}^2)(c^2_{sq}-c^2_{aq})}{(1+w_q-w_q\Omega_m)\left[k^2(1+w_q-w_q\Omega_m) + 9{\cal H}^2\Omega_m(c^2_{sq}-c^2_{aq})\right]}\right]\Phi_d(k) ,\\
V_q &=& \frac{-2}{3{\cal H}(1+w_q-w_q\Omega_m)} \left[\frac{k^2(1+w_q-w_q\Omega_m) + 3\Omega_m(k^2 + 3{\cal H}^2)(c^2_{sq}-c^2_{aq})}{k^2(1+w_q-w_q\Omega_m) + 9{\cal H}^2\Omega_m(c^2_{sq}-c^2_{aq})}\right]\Phi_d(k).
\eea
\end{widetext}
Here $\Phi_d(k)$ is given by \eqref{phid} and
\bea
c^2_{aq} \equiv \frac{p'_q}{\rho'_q} = w_q - \frac{w'_q}{3{\cal H}(1+w_q)} .
\eea
All $z$-dependent quantities above are evaluated at $z_d$, and we assumed $1+w_q(z_d)\neq 0$.

In the \LCDM\, case, $\Delta_q=0=V_q$ and the initial values are
\beq
\Delta_m=-{2k^2 \over 3\Omega_m {\cal H}^2}\Phi_d(k),~~ V_m=-{2 \over 3\Omega_m {\cal H}}\Phi_d(k).
\eeq

%

\end{document}